# STATISTICS AND PROPERTIES OF H II REGIONS IN A SAMPLE OF GRAND DESIGN GALAXIES

## I. LUMINOSITY FUNCTIONS

M. Rozas[1], J.E. Beckman[1], and J.H. Knapen[2]

[1] Instituto de Astrofísica de Canarias, E-38200 La Laguna, Tenerife, Canarias, Spain
[2] Département de Physique, Université de Montréal, C.P. 6128, Succursale Centre Ville, Montréal (Québec), H3C 3J7 Canada

**Abstract.** We present new high quality continuum-subtracted Hα images of the grand-design galaxies NGC 157, NGC 3631, NGC 6764 and NGC 6951. We have determined the positions, angular sizes, and fluxes of their individual H II regions, and describe statistical properties of the H II region samples. We construct luminosity functions for all the H II regions in the disc and separately for arm and interarm zones for each galaxy. The slopes of the luminosity functions for the complete sample agree well with values published for other spiral galaxies of comparable morphological type. For three galaxies we determined the slopes of the luminosity functions for the spiral arm and interarm zones separately. We find that for NGC 157, NGC 3631, and NGC 6951 these slopes are equal within the errors of determination. We compare our results to those found from earlier work, specifically for M51 and NGC 6814, and discuss implications for massive star forming processes.

## 1. Introduction

The distribution of H II regions is an excellent tracer of recent massive star formation in spiral galaxies (see e.g. the review by Kennicutt 1992). Complete catalogues of these regions provide an observational base which is increasingly used in programmes aimed at determining global massive star formation parameters across galactic discs. However, high quality catalogues and detailed statistical analyses are still not frequently found in the literature.

*Send offprint requests to*: J.E. Beckman

This is the first of two papers in which we describe an analysis of the luminosity functions (LFs) and geometrical distribution of H II regions in the discs of NGC 157, NGC 3631, NGC 6764 and NGC 6951. These galaxies were observed as part of our ongoing program to image grand-design spiral galaxies in the Hα emission line.

In the present paper, we will discuss exclusively late-type spiral galaxies, although other galaxies of other morphological types have been studied. For example, for LFs for Sa galaxies we refer to Caldwell et al. (1991). Kennicutt, Edgar & Hodge (1989, hereafter referred to as KEH) have studied the H II region LFs in a sample of 30 galaxies, ranging in morphological type from Sb to Irr. They found that the H II region LF can be described by a power law function, with $N(L) \propto L^{-2\pm0.5}$. Late type spirals possess more H II regions than early type galaxies, and their LF is shallower. KEH constructed arm and interarm LFs for five galaxies in their sample, for which the H II region LF for the interarm is significantly different from that for the arm.

Cepa & Beckman (1989) have studied the distribution of H II regions in the barred spiral NGC 3992, and did not find any significant difference in the arm and interarm LF slopes. The same authors (Cepa & Beckman 1990) published a catalogue of the H II regions in the inner $3' \times 4'$ of NGC 4321. In that galaxy, they *did* find a significant difference between arm and interarm LFs, a result similar to that obtained from an analysis of the H II region LF in M51 by Rand (1992, hereafter referred to as R92). The distribution of H II regions in the disc of NGC 6814 (Knapen et al. 1993, hereafter referred to as KACB) shows that the slopes of the arm and interarm LFs are equal within the uncertainties of the fits to the slopes for that galaxy.

The Hα image used by R92, and the image of NGC 6814 as used by KACB (which forms part of the same program as the galaxies analysed in the present paper) are similar in quality to the images presented now.

Table 1. Basic properties of galaxies.

| Galaxy | Arm Class | Type (RC3) | $R_{25}$ (') | $D$ (Mpc) | R.A. (1950) | Dec (1950) |
|---|---|---|---|---|---|---|
| NGC 157 | 12 | SABbc(rs) | 2.1 | 22.2 | $00^h\,32^m\,14^s.4$ | $-08°\,40'\,20''$ |
| NGC 3631 | 9 | Sc(s) | 2.5 | 15.4 | $11^h\,18^m\,13^s.3$ | $56°\,26'\,43''$ |
| NGC 6764 | 5 | SBbc(s) | 1.2 | 32.2 | $19^h\,07^m\,01^s.5$ | $50°\,51'\,03''$ |
| NGC 6951 | 12 | SABbc(rs) | 1.9 | 19.0 | $20^h\,36^m\,37^s.7$ | $65°\,55'\,48''$ |

We can therefore compare our results directly with those found in R92, KACB, and more qualitatively with KEH.

In Table 1 we give basic properties of the four galaxies presented in this paper: NGC 157, NGC 3631, NGC 6764, and NGC 6951. Column two in the Table is the arm class, as defined by Elmegreen & Elmegreen (1987), where class 12 galaxies have the best-defined spiral arms. Column three is the morphological type (de Vaucouleurs et al. 1991; hereafter RC3). Column four is the radius at 25 mag arcsec$^{-2}$, $R_{25}$ (in arcmin), from the RC3. Column five is the distance (in Mpc), as determined from the redshift, assuming $H_0 = 75$ km s$^{-1}$ Mpc$^{-1}$. Finally, columns six and seven give the positions of the centres of the galaxies for epoch 1950 (from RC3).

The galaxies were selected to give a balance between barred and non-barred objects, with the aim of comparing and contrasting their properties:

- NGC 157 is an angularly small, weakly barred galaxy with well defined spiral arms. The nuclear region is relatively small and it is noted for several dusty regions, which can be clearly recognized on a blue image (not reproduced here).
- NGC 3631 is a late-type spiral galaxy of type Sc. It shows no sign of a bar, and has no obvious companion nearby. It has two main spiral arms and some smaller arm sections, as described by Boeshaar and Hodge (1977), who studied the distribution of H II regions in this galaxy.
- NGC 6764 is an angularly small barred galaxy of arm class 5. It has two spiral arms and very little emission between them. The brightest star forming regions are at the ends of the bar. The nucleus of NGC 6764 is a LINER.
- NGC 6951 is the second mixed-type, weakly barred spiral in our sample. A bright and well defined ring structure surrounds the very bright nuclear region, which is elongated along an axis at nearly right angles to the bar axis. We will not discuss this inner region in the present paper. NGC 6951 hosts a Seyfert nucleus (Boer & Schulz 1993).

In Fig 1 we show grey scale representations of the continuum-subtracted H$\alpha$ images of NGC 157, NGC 3631, NGC 6764, and NGC 6951. The observations and data reduction procedures are summarized in Section 2, and the H II region selection and the construction of the catalogues is described in Sec. 3. The results are presented in terms of LFs in Sec. 4 and briefly discussed in terms of star formation processes in Sec 5. The main conclusions are summarized in Sec 6.

## 2. Observations and data reduction

The observations of NGC 157, NGC 6951 and NGC 6764 were made during the nights of September $7^{th}$ and $8^{th}$, 1990 with the 4.2m William Herschel Telescope (WHT) on La Palma. We used the TAURUS instrument in imaging mode as a re-imaging camera, mounted at the Cassegrain focus of the telescope. The detector used was an EEV CCD 7 with projected pixel size $0''.279 \times 0''.279$. Observing conditions were very good with $0''.8$ seeing (FWHM as measured in the final images) and photometric sky. We obtained two exposures of 1200 seconds each per galaxy: one through a 15 Å wide filter whose central wavelength coincided with that of the redshifted H$\alpha$ emission from the galaxy ($\lambda6601$Å for NGC 157; $\lambda6613$Å for NGC 6764; and $\lambda6589$Å for NGC 6951), and another through a nonredshifted H$\alpha$ filter ($\lambda6565$Å with 15Å width) for continuum subtraction. NGC 3631 was observed during the night of May $20^{th}$, 1992 using the same telescope and instrumentation. Given the angular size of this galaxy, we decided to take three sets of line and continuum images of 1800 s each (composed of 2 separate 900 s images) at three different positions of the galaxy disc. We used the 15Å wide $\lambda6589$Å filter for the line images, and the equally wide $\lambda6565$Å filter for the continuum imaging. H$\alpha$ continuum-subtracted images were produced for these three positions separately (see below) which were then pasted together into one final H$\alpha$ image of the galaxy. The resolution of the final image was limited to the $1''.3$ resolution of the sub-image with the poorest resolution, which also led us to re-bin the final image to a $0''.4 \times 0''.4$ pixelsize. Photometric calibration was achieved by comparison of the final image with an image of the inner part of the disc of NGC 3631 obtained earlier with the WHT.

Standard reduction routines were used, following closely the procedure used in KACB for NGC 6814. A bias level was first subtracted, and the images were then corrected using appropriate dawn sky flatfields. Next, a value for the sky background was obtained by measuring the background and associated noise levels in portions of the images where there was no emission from the galaxies. In order to align the two (line and continuum) images of each galaxy, positions of several foreground stars were determined making gaussian fits, and the off-band image was displaced to coincide with the Hα image using the stellar positions. The accuracy of the superpositions was in all cases better than 0.2 pixels. In the case of NGC 3631, we also used foreground star positions to align the three sub-images for subsequent combination into the final galaxy image.

Fig. 1. Grey scale representations of the Hα continuum-subtracted images of NGC 157 (a); NGC 3631 (b); NGC 6764 (c); and NGC 6951 (d).

Once the aligned images were cleaned of cosmic ray effects (see KACB for a discussion) and an occasional bad pixel, the continuum image was subtracted from the line (Hα + continuum) image, giving nett Hα flux. As we expected (see KACB), the best scaling factor was was close to unity for the subtraction of the continuum image from the line image, given that both images were taken through filters with very similar bandwidths and transmissions and in practically the same observing conditions. We made sure the scaling factor used (1.0) cannot be in error by more than 5%, by experimentally scaling the continuum by factors less than unity ($\leq 0.95$). In that case, residual continuum was left in the nuclear and arm regions after subtraction. If factors greater than unity were employed ($\geq 1.05$), negative intensities appeared in these same regions. As described in KACB, we prefer to use the present method over one where foreground stars (which may not have a flat spectrum in the range considered) are used to scale the images.

Table 2. Calibration constants: Hα luminosity which corresponds to a single instrumental count

| Galaxy | $L_{\mathrm{H}\alpha}$ (erg s$^{-1}$ count$^{-1}$) |
|---|---|
| NGC 157 | $1.47 \times 10^{34}$ |
| NGC 3631 | $0.76 \times 10^{34}$ |
| NGC 6764 | $3.45 \times 10^{34}$ |
| NGC 6951 | $2.08 \times 10^{34}$ |

Absolute flux calibration was carried out using observations of standard stars from the lists of Oke (1974), Stone (1977), and Filippenko & Greenstein (1984). The luminosity in Hα, which corresponds to a single instrumental count in each of our images, is given in Table 2 for each galaxy.

## 3. Production of the H II region catalogues

Table 3. The r.m.s. noise (in instrumental counts; column 2) of the background subtracted Hα images; the lower limit to the luminosity of the detected H II regions in each galaxy (col. 3); and the approximate size of the smallest catalogued regions (col. 4).

| Galaxy | $\sigma$ r.m.s. | $\log L_{\mathrm{H}\alpha}$ (erg s$^{-1}$) | $D$ (pc) |
|---|---|---|---|
| NGC 157 | 3-5 | 36.56 | $\approx 97$ |
| NGC 3631 | 2-6 | 36.42 | $\approx 67$ |
| NGC 6764 | 5-7 | 36.45 | $\approx 140$ |
| NGC 6951 | 2-4 | 36.27 | $\approx 83$ |

Before producing the H II region catalogues, we flagged the foreground stars in the images. These are distinguishable from H II regions by their regular, circular shapes in the original, unsubtracted images, and because they show more intensity in the continuum than in the corresponding Hα continuum-subtracted image. [Ideally, foreground stars should not show up at all in the Hα continuum-subtracted images, but in most cases some residual is seen, due to e.g. differences in point spread function, alignment, or stellar emission between the line and continuum images; or because the star is saturated on one or both images.] Emission in the Hα image coincident with a foreground star on the continuum image was considered residual starlight, and not entered in the catalogue as an H II region. As a selection criterion for H II regions we imposed that a feature must contain at least nine contiguous pixels, each with an intensity of at least three times the r.m.s noise level of the local background. Any object not meeting this criterion was treated as indistinguishable from noise. In Table 3 we list the r.m.s. noise of the background-subtracted Hα images, the lower limit to the luminosities of the detected H II regions in each galaxy, as well as the sizes of the smallest catalogued regions (the last two quantities are derived directly from the adopted selection criterion).

In identifying and quantifying the parameters of the H II regions we had to overcome three complicating effects. Firstly, many H II regions appear to overlap on the image. Without attempting to analyze what fraction of these overlaps implies real contact and what fraction is merely a projection effect, we adopted the solution proposed in R92 and followed in KACB of counting each peak in H$\alpha$ as representing a single H II region. The flux of each H II region was then estimated by integrating over the pixels which could be reasonably attributed to a given region. One will undoubtedly miss a number of H II regions that are too weak to be detected in the vicinity of stronger emitters close by. This will influence the lower end of the LF but is not a significant factor in the determination of the shape of the true LF at the higher luminosity end (R92). Secondly, an H II region is not necessarily circular. We adopted as an effective radius the mean of the maximum and minimum radii as measured for each H II region. Thirdly, the presence of diffuse H$\alpha$ may lead to ill-defined edges of H II regions, introducing some systematic errors, above all for the weakest regions. We have not tried to quantify this effect, but note its possible influence.

After identifying each H II region, we measured the position of its centre with respect to the centre of the galaxy, and derived a mean radius. We determined the flux of each region using a programme which was previously used by KEH, R92 and KACB, that integrates counts within a circular aperture. A constant sky value was subtracted from the images before determining the integrated fluxes (see above), but we made additional corrections for a small number of H II regions due to locally varying background levels. We estimate that sky subtraction errors cannot influence the luminosity determinations by more than about 5% in even the weakest regions.

**Table 4.** Total number of catalogued H II regions for each galaxy, for arm and interarm separately and for the whole galaxy.

| Galaxy   | Arm | Interarm | Total |
|----------|-----|----------|-------|
| NGC 157  | 513 | 195      | 708   |
| NGC 3631 | 902 | 420      | 1322  |
| NGC 6764 | 332 | 16       | 348   |
| NGC 6951 | 603 | 71       | 674   |

Most previous studies of H II regions in spiral galaxies were made using relatively low-resolution imaging. For example, in an early study of H II regions in NGC 3631, Boeshaar & Hodge (1977) identified only 222 regions. Hodge & Kennicutt (1983) identified only 84 regions in NGC 157. We identify 1322 regions in NGC 3631, and 708 in NGC 157, which reflects the improvement in resolution and sensitivity in the present data. There are no previously published H II region catalogues for NGC 6764 or NGC 6951. Table 4 lists the number of catalogued H II regions in each galaxy for the disc as a whole, and for arm and interarm zones separately.

For all the H II regions catalogued for each galaxy, we determined equatorial coordinate offsets from the nucleus and deprojected distances to the centre (in arcsec), using the inclination angles and position angles given by Grosbøl (1985) (NGC 157: $i = 45°$, PA=35°; NGC 3631: 32°, 126°; NGC 6764: 62°, 62°; NGC 6951: 44°, 157°). We also determined the diameter and the H$\alpha$ luminosity (in erg/s) of each H II region, and assigned an index letter to specify whether the given region is in an arm or an interarm zone. The catalogues are available via the CDS, or from the authors. In Fig. 2 we show schematically the positions of the H II regions in the disc of NGC 157, NGC 3631, NGC 6764, and NGC 6951, on a deprojected RA-dec grid centred on the nucleus of each galaxy. Arm and interarm regions are distinguished by using different symbol styles.

## 4. Luminosity functions

### 4.1. Total LFs

The luminosities in H$\alpha$ of the regions catalogued range from some $10^{36}$ to $10^{40}$ erg s$^{-1}$. The regions at the lower end of the luminosity range are in all probability ionized by a single star, so that the LF at these levels essentially traces the stellar LF, although in this range the sample loses its completeness. Most of the regions detected have luminosities between $10^{37}$ and $10^{39}$ erg s$^{-1}$, which corresponds to Lyman continuum luminosities between $10^{49}$ and $10^{51}$ ionizing photons per second. The upper limit to the number of ionizing photons per second from a single star is of order $10^{49}$; thus, the majority of the regions detected must be ionized by the radiation from groups of stars. The brightest regions may be physically different from the others, e.g. formed under different physical conditions, or possibly the result of mergers.

**Table 5.** LF peaks (column 2) and slopes ($\alpha - 1$; col. 3), and lower luminosity limits for completeness (col. 4; $L$ in erg s$^{-1}$).

| Galaxy   | LF peak | LF slope         | completeness limit |
|----------|---------|------------------|--------------------|
| NGC 157  | 37.8    | $-1.89 \pm 0.09$ | $\log L \geq 37.8$ |
| NGC 3631 | 36.4    | $-1.73 \pm 0.08$ | $\log L \geq 37.2$ |
| NGC 6764 | 36.4    | $-1.84 \pm 0.11$ | $\log L \geq 37.6$ |
| NGC 6814 |         | $-2.37 \pm 0.09$ | $\log L \geq 37.6$ |
| NGC 6951 | 36.2    | $-2.21 \pm 0.10$ | $\log L \geq 37.6$ |
| M51      |         | $-1.87 \pm 0.09$ | $log L \geq 37.6$  |

The LFs are presented in Fig. 3, where we have plotted the decimal logarithm of the number of H II regions in bins of width 0.2 dex along the ordinate and the logarithm of the luminosity (in erg s$^{-1}$) along the abscissa. The cutoff levels for each galaxy are a result of our criterion for discriminating H II regions from the background noise; they are visible as a drop in the observed LFs to the low luminosity side, although the LFs in this region are also influenced by incompleteness. The positions of the peaks of the LFs are in good agreement with those determined by R92 for M51 and by KACB for NGC 6814, indicating that the data sets used for these studies and the present one are comparable in quality. In Table 5 we show the peaks of the LFs of the galaxies, as well as the completeness limits, below which faint H II regions, especially in the spiral arm zones, may have been missed. These were defined by departures from uniformity in slope of the observed functions.

**Fig. 2.** Deprojected maps of the program galaxies, composed of schematic representations of the positions of the measured HII regions. Open symbols represent HII regions in the spiral arms, filled symbols HII regions in the interarm disc. R. A. and Dec. offsets are in arcsec, relative to the centre of the galaxy. a. NGC 157 (upper left); b. NGC 3631 (upper right); c. NGC 6764 (lower left); and d. NGC 6951 (lower right).

Omitting the centres of the galaxies and those H II regions below the completeness limits in luminosity, we fitted functions of the form $N(L) = AL^\alpha dL$ to the data. In all cases the slopes lie within the range $\alpha = -2.0 \pm 0.5$, in agreement with KEH. Table 5 lists the slopes of the LFs and the lower luminosity limits of the fits for the four galaxies in the present paper. For comparison we also list NGC 6814 (from KACB) and M51 (from R92). Note that to be consistent with the literature, we add $-1$ to the slopes of the plotted LFs (see KACB).

An interesting characteristic of all the LFs is a "blip" in the otherwise monotonically falling linear function close to $L = 10^{38.6}$ erg s$^{-1}$. We reported this phenomenon in NGC 6814 (KACB), and it is also seen in M 51 (R92). The blip is accompanied by a change in slope, with a steeper gradient at higher luminosities, reported previously by KEH. They discuss this change in terms of a transition from normal to supergiant H II regions, but this is a phenomenological rather than a physical description. The discrepancy in the LFs would then indicate that some well-defined physical mechanism causes the formation of such supergiant H II regions. One possibility is that the "blip" and the change of slope are due to the change from ionization-bounded to density-bounded regions, and marks the limitation on the size of the neutral gas clouds which give birth to the stellar associations. The change of slope shows up very clearly at $\log L = 38.6$ erg s$^{-1}$ in NGC 157 and NGC 6764; in NGC 6951 and NGC 3631 it is just a shade lower: $\log L = 38.5$ erg s$^{-1}$ and $\log L = 38.4$ erg s$^{-1}$, respectively, although in NGC 3631 the break is less sharp than in the others. It will be of great interest to see whether, on theoretical grounds, we should expect the break to occur at constant luminosity, in which case these differences of order 0.1 dex in the luminosity of the break point would correspond to differences in the assumed distances of order 10%.

In Fig. 4 we show the LFs of the galaxies, but now with two fits in two luminosity ranges: above and below the break point in the LF noted above. This change in slope, referred to previously by KEH for some of the galaxies they analyzed, is clearly present in all the galaxies of the present paper. In Table 6 we give the two sets of slopes, and the limits of their respective luminosity ranges. Again we note that the lower range slopes are very close to each other in value, as are also the upper range slopes, implying common underlying physical behaviour from galaxy to galaxy.

### 4.2. Arm and Interarm LFs

In Fig. 5 we show the LFs for the H II regions of the arm and the interarm regions separately, for all the galaxies. Of the galaxies analyzed, NGC 6764 does not show particularly well defined arms in H$\alpha$, which made a classification as interarm or arm H II regions difficult. In all galaxies, we classified as interarm regions those not obviously connected to the nearest spiral arm. We checked the results we obtained this way by comparing the positions of the spiral arms as seen in the continuum image, where the edges of the arms are better defined, and reclassified a few H II regions. In general, the H II regions lying in the interarm zones are far fewer in number than those in the arms, as is evident from Table 4.

The slopes of the LFs of the interarm regions of NGC 157, NGC 3631 and NGC 6951 are equal to those of the arms, within the range of uncertainties inherent in the measurements. The interarm LF for NGC 6764 has no statistical significance, since the number of interarm regions is very small.

The slopes of the LFs for the arms and the interarm zones for each galaxy, and the limits to the luminosities for which they were determined, appear in Table 7. The table also incorporates NGC 6814 (KACB), and M51 (R92). Due to the smaller numbers of H II regions, the LFs of the interarm regions are displaced vertically with respect to the arm LFs, and the peaks are displaced horizontally, to lower luminosities.

## 5. Implications for star formation

The most significant result of this part of the study of the distribution of H II regions in the discs of NGC 157, NGC 3631, NGC 6764, and NGC 6951 is the determination of slopes of the LFs not only in the disc as a whole, but also in the arms and the interarms zones separately. Within the errors of determination, we find that the arm and interarm LFs have the same slopes for NGC 157, NGC 3631, and NGC 6951. For NGC 6764 there is little

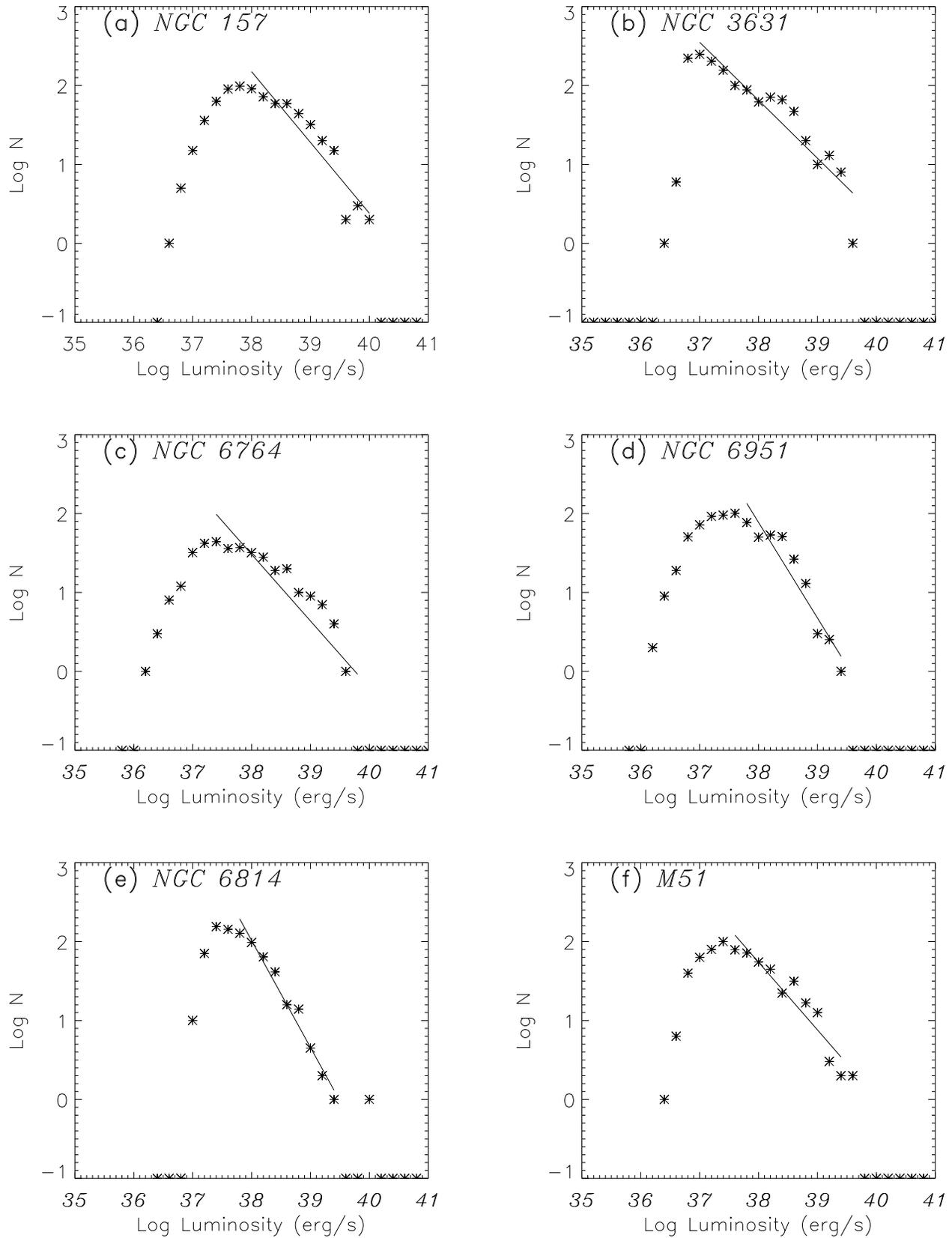

**Fig. 3.** HII region LFs of NGC 157 (a, upper left); NGC 3631 (b, upper right); NGC 6764 (c, middle left); NGC 6951 (d, middle right); NGC 6814 (e, lower left; from KACB); and M51 (f, lower right; from R92). Lines indicate best fits (see text)

**Table 6.** LF slopes and luminosity limits for the two luminosity ranges in erg s$^{-1}$.

| NGC 157 | | NGC 3631 | | NGC 6764 | | NGC 6951 | |
|---|---|---|---|---|---|---|---|
| range | slope | range | slope | range | slope | range | slope |
| $37.8 \leq logL \leq 38.6$ | $-1.31$ | $37.0 \leq logL \leq 38.2$ | $-1.51$ | $37.6 \leq logL \leq 38.6$ | $-1.30$ | $37.6 \leq logL \leq 38.6$ | $-1.33$ |
| $38.6 \leq logL \leq 40.0$ | $-2.18$ | $38.2 \leq logL \leq 39.6$ | $-2.16$ | $38.6 \leq logL \leq 39.6$ | $-2.11$ | $38.6 \leq logL \leq 39.4$ | $-2.74$ |

**Table 7.** Slopes of arm and interarm LFs and corresponding lower luminosity limits in erg s$^{-1}$.

| Galaxy | Arm | | Interarm | |
|---|---|---|---|---|
| | slope of LF | Luminosity limit | slope of LF | Luminosity limit |
| NGC 157 | $-1.85 \pm 0.9$ | $\log L \geq 37.8$ | $-2.00 \pm 0.19$ | $\log L \geq 36.6$ |
| NGC 3631 | $-1.61 \pm 0.10$ | $\log L \geq 37.2$ | $-1.78 \pm 0.12$ | $\log L \geq 37.0$ |
| NGC 6764 | $-1.80 \pm 0.11$ | $\log L \geq 37.6$ | $-1.39 \pm 0.32$ | $\log L \geq 37.8$ |
| NGC 6814 | $-2.19 \pm 0.08$ | $\log L \geq 37.6$ | $-2.26 \pm 0.19$ | $\log L \geq 37.2$ |
| NGC 6951 | $-1.93 \pm 0.10$ | $logL \geq 37.6$ | $-1.72 \pm 0.20$ | $logL \geq 37.0$ |
| M51 | $-1.82 \pm 0.09$ | $\log L \geq 37.6$ | $-2.05 \pm 0.15$ | $\log L \geq 37.4$ |

statistical significance to the slope of the interarm function.

The spiral arms of NGC 157, NGC 6951, and NGC 3631 are all well defined. The majority of the H II regions are located in the arms, as are *all* the most luminous regions. KEH measured the H II regions of a set of galaxies with well defined spiral arms, separating the arm from the interarm zones, and found as a general result that the slopes are different, due to the presence of a larger number of high-luminosity H II regions in the arms than in the interarm zones. This result is, however, not necessarily reliable, since the effect of H II region crowding in the arms may have affected the luminosity limit in the photographic data of KEH (which have relatively poor resolution and S:N ratio), more so than in the more recent studies (including the present one), and especially in the arms.

Cepa & Beckman (1989) found that the slopes of the arm and interarm LFs are the same for NGC 3992, but did find a significant difference in NGC 4321 (Cepa & Beckman 1990). Rand (1992) attributed the difference he observed in M51 between arm and interarm LF to a less steep mass spectrum of molecular clouds in the arms, i.e. to a higher proportion of massive clouds, giving rise to relatively more H II regions of higher luminosity. It is reasonable to assume that the coalescence of clouds in the arms is due to the enhanced global gas density and to the greater frequency of cloud-cloud interactions there.

Finally, KACB found that in NGC 6814 the slopes of the arm and interarm LFs are the same, within the uncertainties. The authors stated that in NGC 6814 the spiral arms might not be strong enough to build up the large cloud masses needed to produce the number of giant HII regions that make the arm LF shallower than the interarm LF. This could well be the reason that the slopes of the arm and interarm LFs in NGC 157, NGC 3631, and NGC 6951 are the same.

Comparing the results of this paper with those published in the literature, we see that in several grand-design spirals the slope of the arm LFs are shallower than the inter-arm LFs, but that this is certainly not a uniform picture. It is interesting to note that for the three galaxies in this paper for which we were able to determine arm and interarm LF slopes, we cannot distinguish these slopes within the errors of determination. Comparing only the more recently published data of good and comparable quality, we find that only M51 (R92) has significantly differing arm and interarm LF slopes, whereas in NGC 6814 (KACB), NGC 157, NGC 3631 and NGC 6951 (this paper) one cannot distinguish arm from interarm LF slopes. M51 is a galaxy with an exceptionally well-developed and strong spiral system, where the high gas surface densities combined with strong density wave compression in the arms may lead to a shallower molecular cloud mass spectrum, and a shallower arm H II region LF (as suggested by R92). In the other galaxies the arms may not be strong enough to make the LF there differ significantly from the LF in the interarm discs.

## 6. Conclusions

The main results of the present paper, where we analyze the LFs of the H II regions in the discs of NGC 157, NGC 3631, NGC 6764, and NGC 6951, are summarized below.

1. From high quality continuum-subtracted H$\alpha$ images of the grand-design spirals NGC 157, NGC 3631, NGC 6764, and NGC 6951, we have catalogued a total of 708, 1322, 348, and 674 H II regions, respectively. The catalogues include positions, radii and fluxes of all

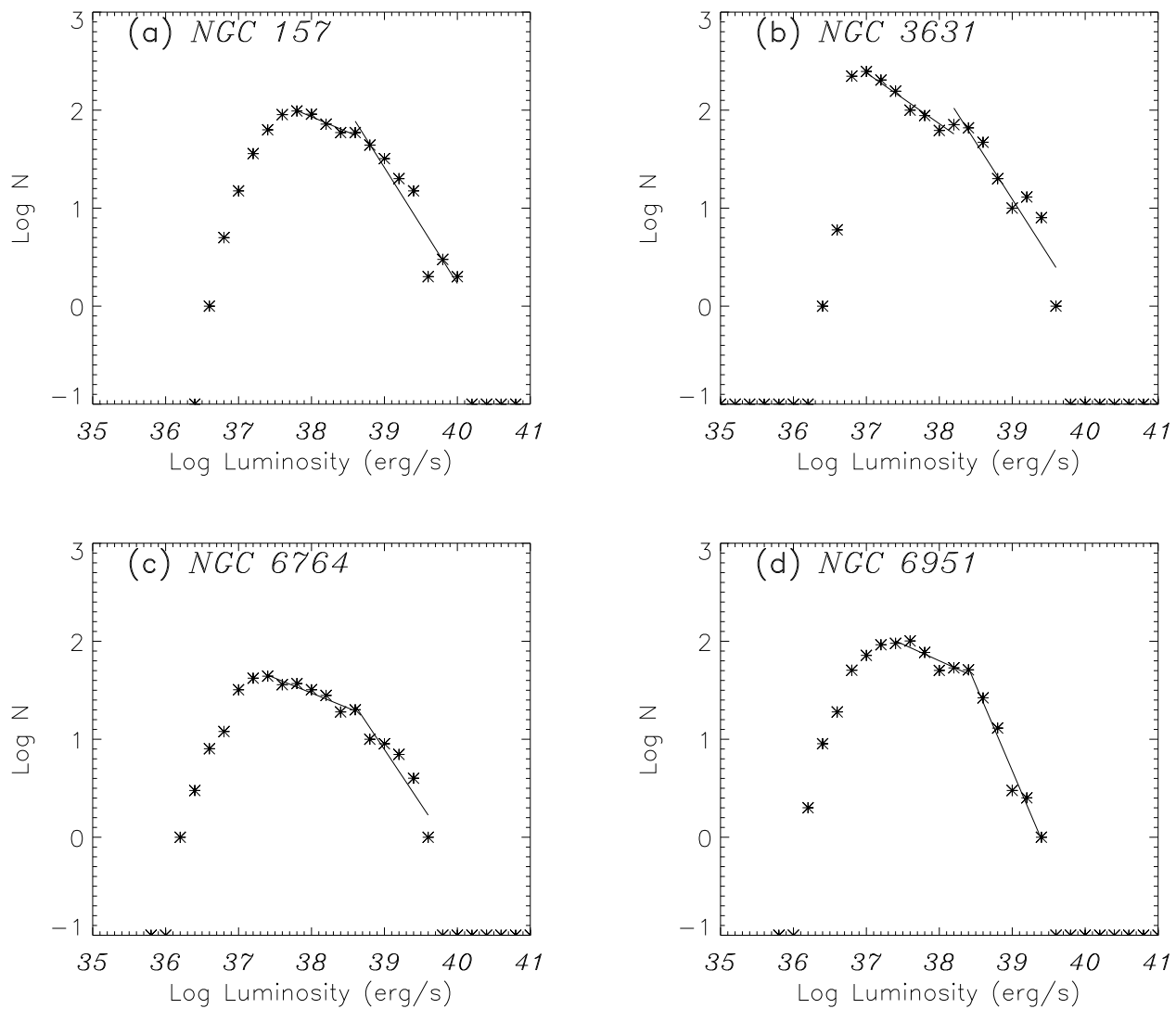

**Fig. 4.** HII region LFs of NGC 157 (a, upper left); NGC 3631 (b, upper right); NGC 6764 (c, lower left); and NGC 6951 (d, lower right). Lines indicate best fits for two luminosity ranges. The slopes are clearly different for the lower and upper luminosity ranges

H II regions. Tables containing all these data are available through CDS or directly from the authors.
2. The LFs for the whole discs ("total LFs") have slopes that agree well with slopes for other galaxies of comparable morphological types.
3. The LFs for the different subsamples of H II regions have the same slopes: in the arms and in the interarm regions, within the limits of the measurement uncertainties.
4. Comparing the LF slopes for arm and interarm regions as determined for NGC 157, NGC 3631, NGC 6764, and NGC 6951, with results for other grand-design galaxies in the literature (especially M51 and NGC 6814), we find that they are different in some of these galaxies, but equal in others. This difference may be due to the presence of sufficiently strong arms to change the arm LF in those grand-design galaxies, such as M51, with strong arms, where coherent density wave effects are of importance.

*Acknowledgements.* We acknowledge Mr. L. Gentet for his extensive help in the production of the H II region catalogue for NGC 3631. We thank Dr. A.F.J. Moffat and Dr. R.F. Peletier for helpful comments on the manuscript. The William Herschel Telescope is operated on the island of La Palma by the Royal Greenwich Observatory in the Spanish Observatorio del Roque de los Muchachos of the Instituto de Astrofísica de Canarias. This work was partially supported by the Spanish DG-ICYT (Dirección General de Investigación Científica y Téc-nica) Grant No. PB91-0525. This research has made use of the NASA/IPAC Extragalactic Database (NED) which is operated by the Jet Propulsion Laboratory, California Institute of Technology, under contract with the National Aeronautics and Space Administration.

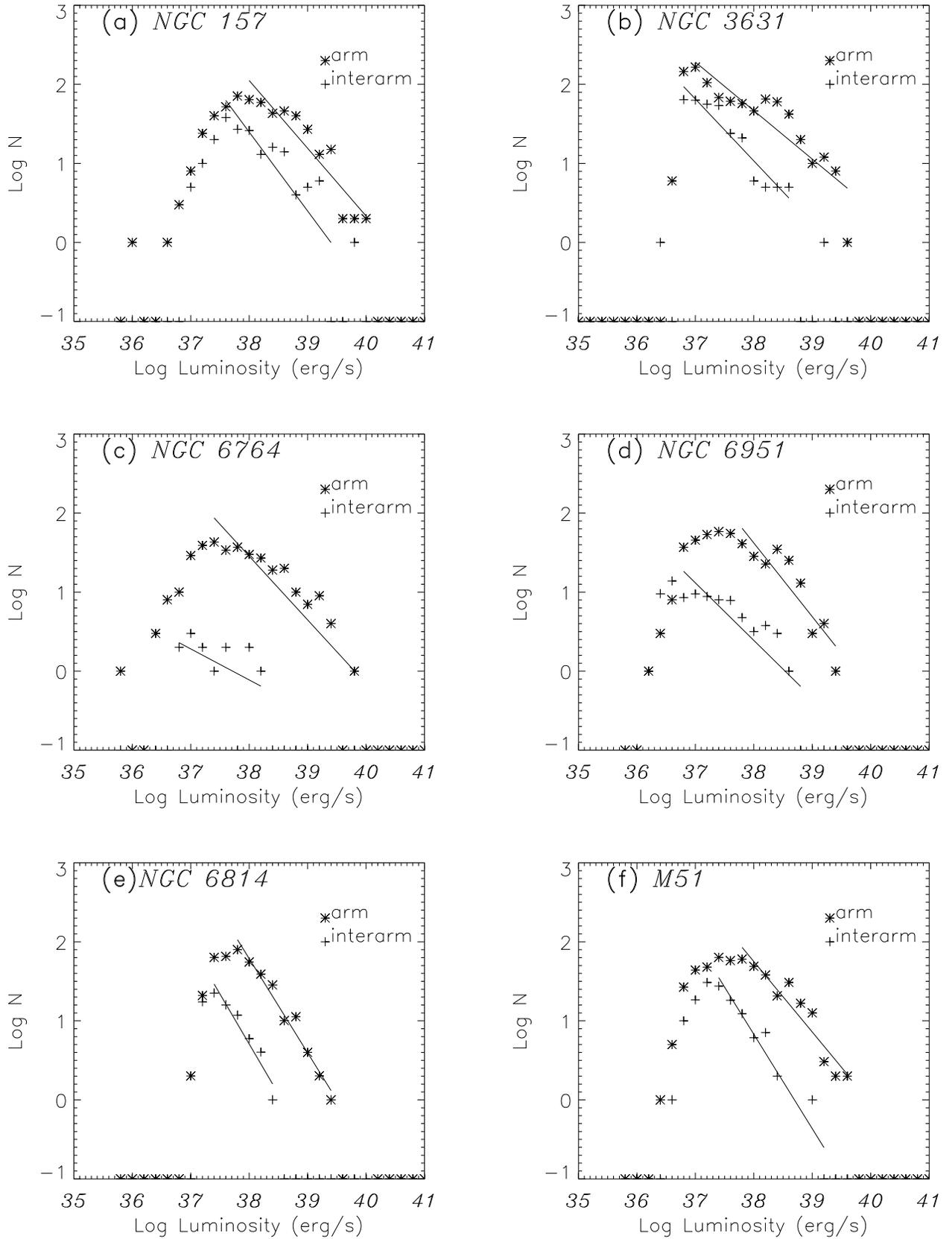

**Fig. 5.** Arm and interarm H II region LFs of NGC 157 (a); NGC 3631 (b); NGC 6764 (c); NGC 6951 (d); NGC 6814 (e, from KACB); and M51 (f, from R92). Drawn lines indicate best fits